\documentclass{optica-article}

\journal{oe}

\usepackage{graphicx}
\usepackage{dcolumn}
\usepackage{bm}
\usepackage{comment}
\usepackage{textcomp}
\usepackage{ulem}
\usepackage[]{graphicx,xcolor}
\usepackage{svg}
\usepackage{amsmath}
\usepackage{bbold}
\usepackage[utf8]{inputenc}
\usepackage[T1]{fontenc}
\usepackage[english]{babel}
\usepackage{verbatim}
\usepackage{textcomp}
\usepackage{ulem}
\usepackage[]{graphicx,xcolor}
\usepackage{soul}

\begin{document}

\title{Optimization of laser stabilization via self-injection locking to a whispering-gallery-mode microresonator: experimental study}

\author{Artem E. Shitikov,\authormark{1,*} Ilya I. Lykov,\authormark{1} Oleg V. Benderov,\authormark{2} Dmitry~A.~Chermoshentsev,\authormark{1,2,3} Ilya K. Gorelov,\authormark{1,4} Andrey N. Danilin,\authormark{1,4} Ramzil R. Galiev,\authormark{5} Nikita M. Kondratiev,\authormark{5} Steevy J. Cordette,\authormark{5} Alexander V. Rodin,\authormark{2} Anatoly V. Masalov,\authormark{1,6} Valery E. Lobanov,\authormark{1} Igor A. Bilenko,\authormark{1,4}}

\address{\authormark{1}Russian Quantum Center, 143026 Skolkovo, Russia\\
\authormark{2}Moscow Institute of Physics and Technology, 141701, Dolgoprudny, Russia\\
\authormark{3}Skolkovo Institute of Science and Technology, Moscow 143025, Russia\\
\authormark{4}Faculty of Physics, Lomonosov Moscow State University, 119991 Moscow, Russia\\
\authormark{5}Directed Energy Research Centre, Technology Innovation Institute, Abu Dhabi, United Arab Emirates\\
\authormark{6}Lebedev Physical Institute, Russian Academy of Sciences, 119991, Moscow, Russia\\
}

\email{\authormark{*}Shartev@gmail.com} 



\begin{abstract}
Self-injection locking of a diode laser to a high-quality-factor microresonator is widely used for frequency stabilization and linewidth narrowing. We constructed several microresonator-based laser sources with measured instantaneous linewidths of 1 Hz and used them for investigation and implementation of the self-injection locking effect. We studied analytically and experimentally the dependence of the stabilization coefficient on tunable parameters such as locking phase and coupling rate. It was shown that precise control of the locking phase allows fine tuning of the generated frequency from the stabilized laser diode. We also showed that it is possible for such laser sources to realize fast continuous and linear frequency modulation by injection current tuning inside the self-injection locking regime. We conceptually demonstrate coherent frequency-modulated continuous wave LIDAR over a distance of 10 km using such a microresonator-stabilized laser diode in the frequency-chirping regime and measure velocities as low as sub-micrometer per second in the unmodulated case. These results could be of interest for cutting-edge technology applications such as space debris monitoring and long-range object classification, high resolution spectroscopy and others.
\end{abstract}

\section{Introduction}


The development of narrow-linewidth and highly stable lasers is one of the key tasks of cutting-edge technologies. Such lasers provide unique opportunities in variety of progressive areas as coherent communication \cite{Fulop2018, Marin-Palomo2017}, ultrafast optical ranging \cite{Trocha2018, Suh2018, Riemensberger2020},  atomic clocks \cite{Newman2019}, astronomy \cite{Suh2019}, and others. Narrow-linewidth lasers are used in coherent Doppler LIDARs for aircraft wake  vortex measurements \cite{wu2019aircraft}, aerosol detection \cite{dai2021calibration} and remote spectroscopic measurements \cite{wang20201645}. Highly stable single frequency lasers are critically important in inverse synthetic aperture LADAR systems (ISAL). ISAL is actively used for space debris monitoring and controlling \cite{mo20183}, long distance object classification with spatial resolution far beyond diffraction limit of the receiving aperture \cite{wang2018inverse}.   

Self-injection locking (SIL) of a diode laser frequency to an eigenfrequency of a high-Q whispering gallery mode (WGM) microresonator provides outstanding results in laser spectral characteristics enhancement \cite{galiev2018spectrum}. It proved to be a simple and robust way of laser stabilization. Since it was demonstrated for the first time \cite{VASSILIEV1998305} SIL still attracts increasing interest and still evolving. The comprehensive theory of the SIL was developed in \cite{Kondratiev2017} and optimal regimes of laser stabilization were discussed in \cite{Galiev20}.  Laser stabilization to sub-Hz linewidth was demonstrated with crystalline microresonator in \cite{liang2015ultralow}, and with on-chip microresonator in \cite{Li2021}. Recently, simultaneous stabilization of two diode lasers by one microresonator has been studied theoretically and experimentally \cite{chermoshentsev2022dual}. 
Furthermore, it was shown that SIL phenomena could be efficiently used to demonstrate the formation of bright \cite{Pavlov2018,  Raja2019, Kondratiev2020, Li2021, Voloshin2021} and dark \cite{Lihachev2022,Jin2021} microcomb solitons, ultra-low-noise photonic microwave oscillators \cite{Liang2015v1}, frequency-modulated continuous wave LIDARs \cite{lihachev2022low}. 
The advantages of using SIL phenomena of semiconductor laser diodes with integrated microresonators are the possibility of on-chip realization \cite{xiang2021laser} of such system that makes the technology compact and inexpensive. Lasers based on the self-injection locking of laser diodes to integrated microresonators from silicon nitride demonstrate outstanding performance \cite{Jin2021, guo2022chip, ling2022self}. Nevertheless, crystalline microresonator-based SIL lasers provides better phase noise \cite{Liang2015, Liang2015v1} and greater opportunities for research and optimization of the key parameters of the effect due to the flexibility unattainable for integrated systems.

In this work we demonstrate comprehensive study of self-injection locking phenomenon, with accurate tunability of different experimental parameters and controllable switching between different SIL regimes. We discuss the possibility of adaptation of the parameters of the self-injection locking scheme for various applications. We studied the influence of the phase shift of the backscattered wave (locking phase) and laser-to-microresonator coupling efficiency (loading) on the performance of the self-injection locked laser (the spectral characteristics of the resulting radiation, the stabilization coefficient, the width of the locking range, and the resulting laser frequency) and found out interesting opportunities for some up-to-date applications.

For detailed experimental SIL investigation we assembled several experimental setups  with precise translation stages to vary key parameters including two fully-packaged turn-key SIL diode lasers. The beatnote signal of two laser diodes stabilized by high-Q MgF$_2$ crystalline microresonators demonstrated an instantaneous 1 Hz linewidth. 
Special attention was paid to the dependence of the SIL parameters on the locking phase defined by the optical path between laser and microresonator. Using spectrogams, we visualized experimental tuning curves for different locking phases. Varying the locking phase, we found out the possibility of the fast fine-tuning of laser diode generation frequency.
Also, it was revealed that, at particular value of the locking phase for high-Q WGMs, the tuning curve splitting can be achieved. 

Another parameter, that can be precisely controlled by the adjustment of the gap between the coupler and the microresonator, is the coupling rate. We analyzed experimental dependence of the stabilization coefficient and showed that its maximum does not coincide with critical coupling, unlike locking width. The experimental results are in good agreement with ours theoretical predictions \cite{Kondratiev:17}.    

  
  

Beside frequency modulated continuous wave (FMCW) LIDARs, continuous tunability of a laser frequency is very attractive for tunable diode laser absorption spectroscopy (TDLAS)\cite{zhang2021portable} and laser cooling. 
Studying tuning curves in the self-injection locking regime, we revealed that there is an area where the frequency changes linearly with driving current. We demonstrated that the frequency tuning inside locking range can be realized up to dozens MHz without dramatic linewidth degradation. We showed that this fact can be used to realize linear frequency modulation up to 200 kHz and the amplitude of the frequency chirp can be controlled by the locking phase. Tuning the frequency of a laser in the self-injection locking regime in a linear way without using an additional acousto-optic modulator seems to be an extremely attractive opportunity for the implementation of devices such as LIDARs, spectrographs, and optical sensors. We experimentally demonstrated the possibility of coherent detection of a self-heterodyne signal through a delay line of 10 km using such frequency modulation (laser chirping) in the self-injection locking regime.

\section{Stabilization coefficient control via loading and locking phase: theory}

The stabilization coefficient $K$ is determined as the inverse derivative (slope) of the generation frequency over the free laser frequency (laser cavity):
\begin {equation}
K = \left(\frac {\delta\omega_{\rm gen}}{\delta\omega_{\rm free}}\right)^{-1}=\frac {K_{\rm free}}{K_{\rm lock}},
\label {K}
\end {equation}
where $\delta\omega_{\rm free}$ is the frequency variation in the free-running regime, $\delta\omega_{\rm gen}$ is the generation frequency variation, which is small in the locked regime providing high $K$. In experiment it is convenient to present the stabilization coefficient as a ratio of the free-running laser frequency change with the injection current $K_{\rm free}$ in unlocked regime to the locked laser frequency change $K_{\rm lock}$.
The stabilization coefficient $K$ determines the linewidth of the locked laser \cite {Laurent1989,Kondratiev:17}:

\begin {equation}
\delta\nu_{\rm locked} = \frac {\delta\nu_{\rm free}}{K^2},
\label {K2}
\end {equation}
where $\delta\nu_{\rm locked}$ and $\delta\nu_{\rm free}$ are the linewidths of the locked and free-running laser, respectively.

We used the expression from \cite{Galiev20} for the stabilization coefficient as a function of five SIL parameters. We suppose the detuning of the laser frequency $\zeta = 2(\omega-\omega_m)/\kappa_m$ from the WGM eigenfrequency close to zero, since we consider the case of optimal stabilization \cite{Galiev20}. Here $\kappa_{\rm m} = \kappa_{\rm mc} + \kappa_{\rm mi}$ is the microresonator's mode decay rate with $\kappa_{\rm mi}$ determined by the intrinsic losses and $\kappa_{\rm mc}$ by the coupling; $\omega$ and $\omega_m$ are the pump output frequency and eigenfrequency of the microresonator. In this case, one can obtain the following expression for the stabilization coefficient:
\begin{equation}
K = 1 + 4\frac{\bar\kappa_{\rm do}}{\kappa_{\rm mi}}\eta(1-\eta)\beta\frac{(2+\frac{\kappa_{\rm m}\tau_s}{2}(\beta^2+1))\text{cos}\psi}{(\beta^2+1)^2},
\label{StabCoeffGeneral}
\end{equation}
where $\bar{\kappa}_{\rm do}$ is the laser effective output beam coupling rate. The corresponding values of the loaded and intrinsic quality factors are defined as $Q_{\rm m} = \omega/\kappa_{\rm m}$, $Q_{\rm int} = \omega/\kappa_{\rm mi}$ and $\eta = \kappa_{\rm mc}/\kappa_{\rm m}$ is the coupling coefficient characterising microresonator loading, $\beta = 2\gamma/\kappa_{\rm m}$ is the normalized mode-splitting coefficient  \cite{gorodetsky2000rayleigh}, $\gamma$ is the forward-backward wave coupling rate, $\tau_s$ is the feedback round-trip time.

\begin{figure}[htbp!]
  \centering
  \includegraphics[width=0.8\linewidth]{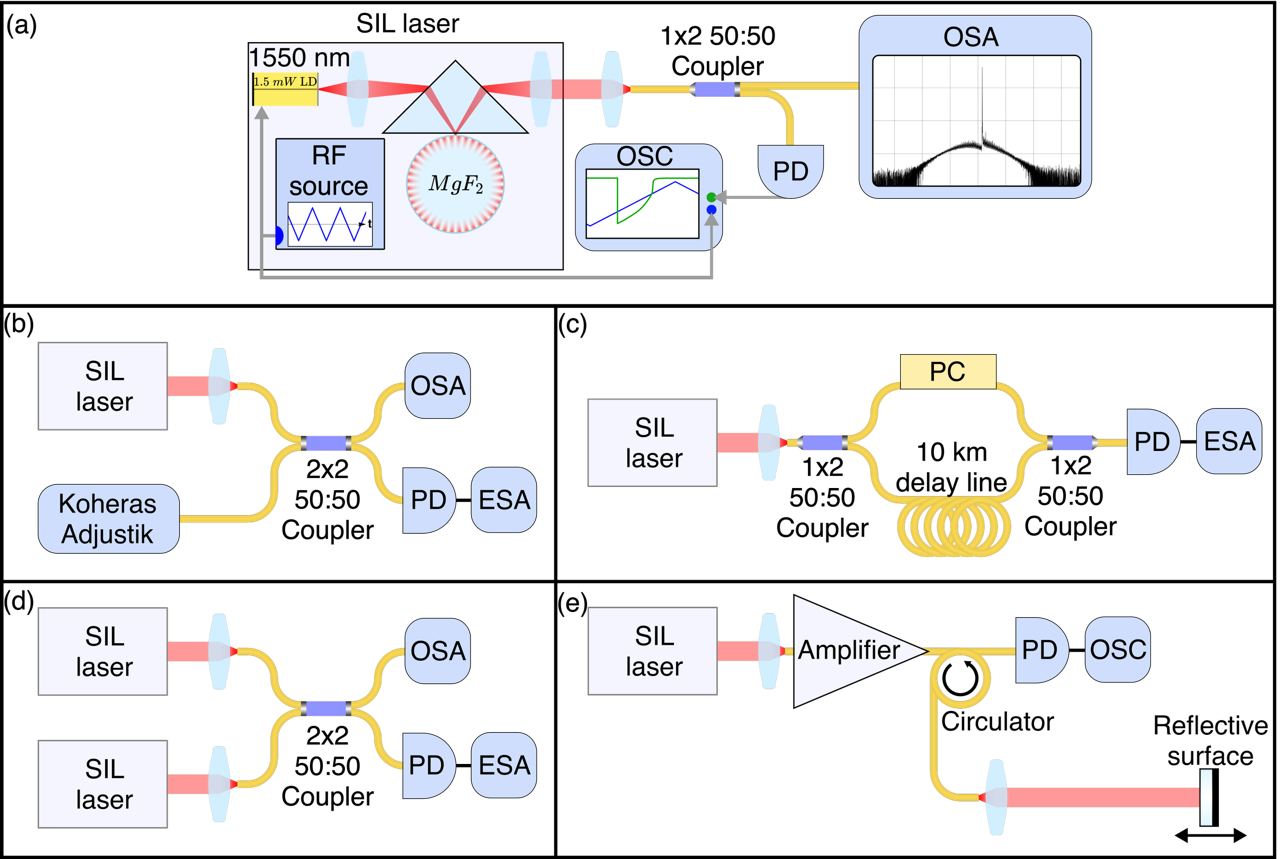}
  \caption{Sketches of the experimental setups: (a) General optical scheme, which includes block "SIL laser" consisting of a laser diode without isolator exciting WGMs through a coupling prism, and RF source that varies laser injection current. That "SIL laser" block presents in the following setups. (b) Heterodyning of SIL laser with reference laser to figure out the dependence from the locking phase. (c) Self-heterodyning linewidth measurement scheme with delay line. (d) Heterodyning of two different SIL lasers. (e) Reflector velocity measurement setup with SIL laser. Here ESA is the electrical spectrum analyzer, OSA is the optical spectrum analyzer, OSC is the oscilloscope, PD is the $\rm 5\ GHz$ bandwidth photodetector, RF source is the source for LD injection current modulation.}
  \label{fig:Setup}
\end{figure}

Let us take the following equation from \cite{shitikov2020microresonator} for the critical coupling regime ($\kappa_{\rm mi} = \kappa_{\rm mc}$):
\begin{equation}
\bar\kappa_{\rm do} = \frac{4}{3\sqrt{3}}\delta\omega_{\rm crit}\frac{\kappa_{\rm mi}}{\gamma},
\label{Kappa_do}
\end{equation}
where $\delta\omega_{\rm crit}$ is the width of the locking range of the stabilized laser in the critical coupling regime. This equation can be substituted in \eqref{StabCoeffGeneral}. After that considering the limit of small $\beta\leq0.1$, that is exact true for high-Q crystalline microresonators in the most cases, one can simplify \eqref{StabCoeffGeneral}:
\begin{equation}
K = 1 + \frac{64}{3\sqrt{3}}\delta\omega_{\rm crit}\frac{\kappa_{\rm mc}\kappa_{\rm mi}}{(\kappa_{\rm mi}+\kappa_{\rm mc})^3}\text{cos}\psi.
\label{StabCoeffSimplified}
\end{equation}
If critical coupling is assumed, then \eqref{StabCoeffSimplified} will take the following form:
\begin{equation}
K =  1 + \frac{8}{3\sqrt{3}}\frac{\delta\omega_{\rm crit}}{\kappa_{\rm mi}}\text{cos}\psi.
\label{StabCoeffCritical}
\end{equation}

That equation directly bounds stabilization coefficient and locking phase, the parameters that can be measured in experiment.

In Eq. \eqref{StabCoeffSimplified} $\kappa_{\rm mc}$ depends on the gap between the coupling element (prism) and the microresonator as follows \cite{gorodetsky1994high}:
\begin{equation}
\kappa_{\rm mc}\approx\frac{\omega}{2}\dfrac{{\rm exp}{\{-2kd\sqrt{n^{2}-1}\}}}{(\dfrac{n^{2}-1}{n}ak)^{3/2}}\times
  \begin{cases}
   \frac{1}{\sqrt{\pi/(1+\sqrt{n^{2}-1})}}       \text{,}\; p = 0\\
    1/(\sqrt{2 p}\pi)  \;\;\;\;\;\;\;  \text{,}\; p > 0
  \end{cases}
\label{kappa}
\end{equation}
where $a$ is the radius of a microresonator, $\omega$ is the optical pump frequency, $k = \omega/c$ is the wavenumber, $p\in\mathbb{N}$ is the WGM vertical index, $d$ is the prism-microresonator gap; this equation is valid in case of close values of refractive index (n) of microresonator and coupling element. It is worth noting that this equation was obtained for the case of the spherical microresonator, but it was shown that for the spheroid microresonator the given equation can be used with great accuracy  \cite{Coupling18, shitikov2020microresonator}. Substituting Eq. \eqref{kappa} into Eq. \eqref{StabCoeffSimplified}, we then obtain the stabilization coefficient as a function of $d$.


Thus, one can see from obtained analytical formulas that spectral characteristics of the laser sources stabilized via SIL effect depends on the locking phase value and on loading defined by the prism-microresonator gap. Further, these dependencies will be checked experimentally.

\section{Experimental setups}


A 1550 nm distributed feedback (DFB) laser with power up to $1.5$ mW is used to excite the WGMs in bulk crystalline MgF$_2$ microresonator. The injection current of the laser diode can be controlled by an external generator. A schematic sketch of the experimental setup is shown in Fig.~\ref{fig:Setup}(a). The laser with a lens is installed on a high-precision translation stage with PZT, which allows us to control the distance between the laser and the resonator cavity with an accuracy of tens of nanometers. The resonator is also set to a translation stage with PZT with a step of 27 nm to control the distance between the resonator and the coupling prism. Laser injection current is controlled with RF generator. These elements (laser diode, RF generator, coupling prism, microresonator and lenses) compose "SIL laser" block. This block is also used as a component of the following setups. After the coupling element, the laser beam is coupled into a single-mode optical fiber and proceeded to measurement equipment. 

First, we characterized the self-injection locked laser diode by beating its output signal with the reference laser (Koheras Adjustik with instantaneous linewidth of less than 100 Hz) providing heterodyne scheme [see Fig.~ \ref{fig:Setup}(b)]. The wavelength of the reference laser is matched to the wavelength of the laser under test to achieve microwave beatnote. The result beatnote signal was analyzed with a 5 GHz broadband detector, which signal was observed at an electrical spectrum analyzer (ESA) and an oscilloscope. This setup was used to visualize tuning curves for different locking phases.


We make noise characterization of the self-injection locked laser diodes.  In case of the absence of laser much narrower than the laser under study, the optimal way is to heterodyne it with an equivalent laser. Fitting the obtained beatnote with a Voigt profile one may estimate the laser linewidth. We measured phase noise spectral density of the beatnote of two SIL laser diodes by heterodying it [see Fig.~\ref{fig:Setup}(d)]. Signal was measured with electrical spectrum analyzer with an option of quadrature analysis through detector with bandwidth of 45 GHz.

Fully characterized lasers are supposed to be ideal candidates for some proof-of-concept experiments. One of it is to measure a delay line length using linearly modulated laser in SIL regime [see Fig.~\ref{fig:Setup}(c)]. The laser radiation is divided into two parts, the first goes through an optical fiber about 40 cm long with a fiber polarization controller in it, the second along a 10 km long delay line, then the 2 radiation parts are then summed through another fiber beam splitter and fed into a detector. The signal from the fiber detector can be examined with an ESA and an oscilloscope. We were able to use a frequency modulated continuous wave scheme by modulating the laser injection current and consequently modulate output frequency inside SIL regime. The frequency modulation parameters can be manipulated by varying locking phase. 

Another application is a velocity sensor based on a Doppler frequency shift [see Fig.~\ref{fig:Setup}(e)]. SIL laser diode beam is amplified with an optical amplifier up to 30 mW. Then light through a circulator goes to a free space where reflects from the target with diffusion surface. The target is mounted on the piezoelectric stage which is continuously moving with low constant velocity of several $\mu$m per second. That velocity can be measured by detecting a Doppler-shifted frequency.

\section {Locking phase tuning}

We performed experimental investigation of the modification of SIL dynamics and stabilized laser parameters upon locking phase tuning.

One may change locking phase varying the distance between the laser diode and microresonator. The injection current of the diode laser was swept to obtain generation frequency in a vicinity of the WGM eigenmode, so that self-injection locking took place. Calculating spectrograms of the beatnote signal allows to visualize a tuning curve for different locking phases and provides detailed information about the self-injection locking dynamics. A spectrogram is a two-dimensional representation of the dependence of the power spectral density of the signal on time, with the signal intensity represented in color. The spectrograms are calculated using the window Fourier transform with the Blackman window.

\begin{figure}[htbp!]
\centering
\includegraphics[width= \linewidth]{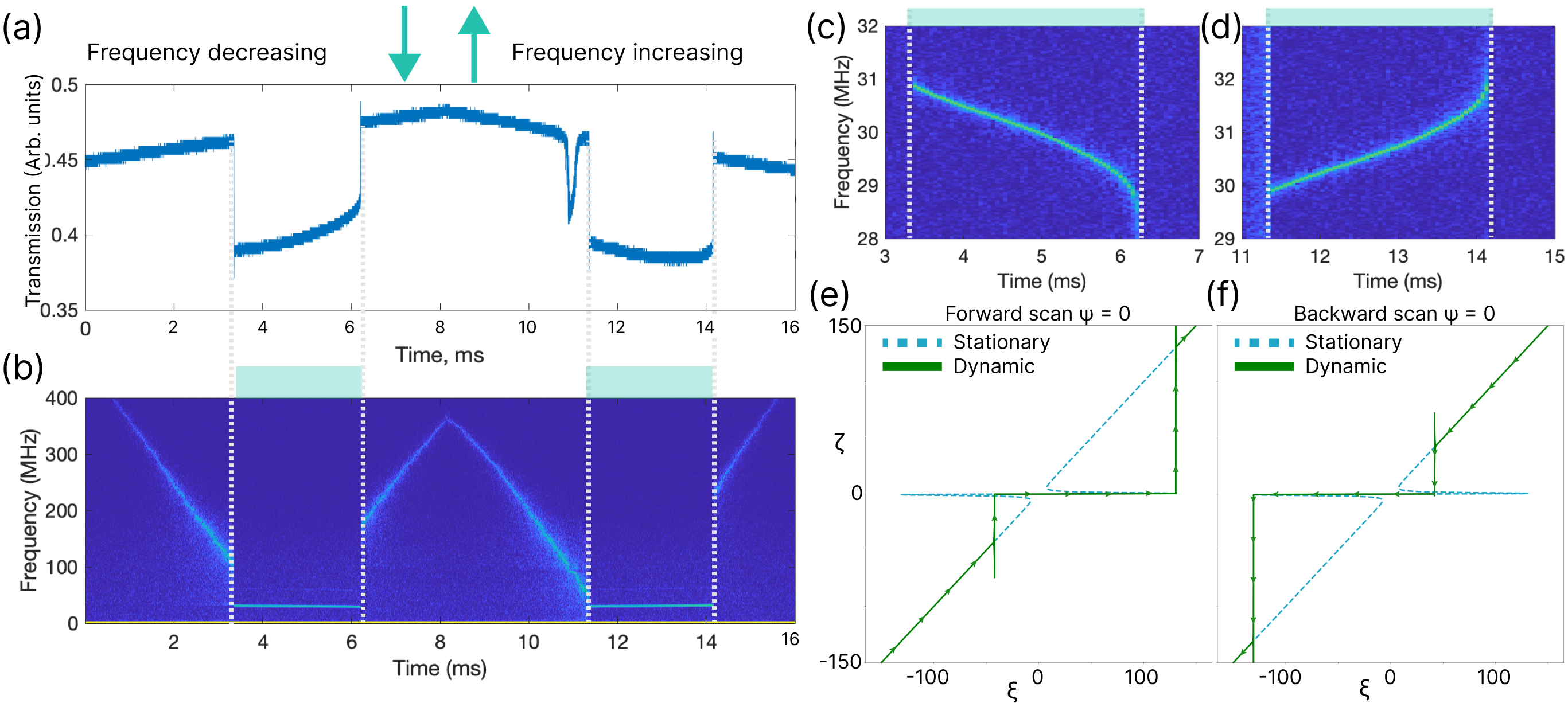}

\caption{SIL dynamics at locking phase $\psi = 0$. The transmission signal is presented in the top graph (a). High-order mode appeared at $\approx$ 11 ms. The spectrogram of the beatnote signal with the reference laser is presented in the middle (b). Enlarged areas of the spectrograms  corresponding to locking ranges are presented in the panels (c, d). The modeling results are presented in the bottom panels (e, f). Analytical stationary solutions are plotted with blue dashed line and dynamic modeling with solid green line.
}

\label{fig:PhaseComparison0}

\end{figure}

\begin{figure}[htbp!]
\centering
\includegraphics[width=\linewidth]{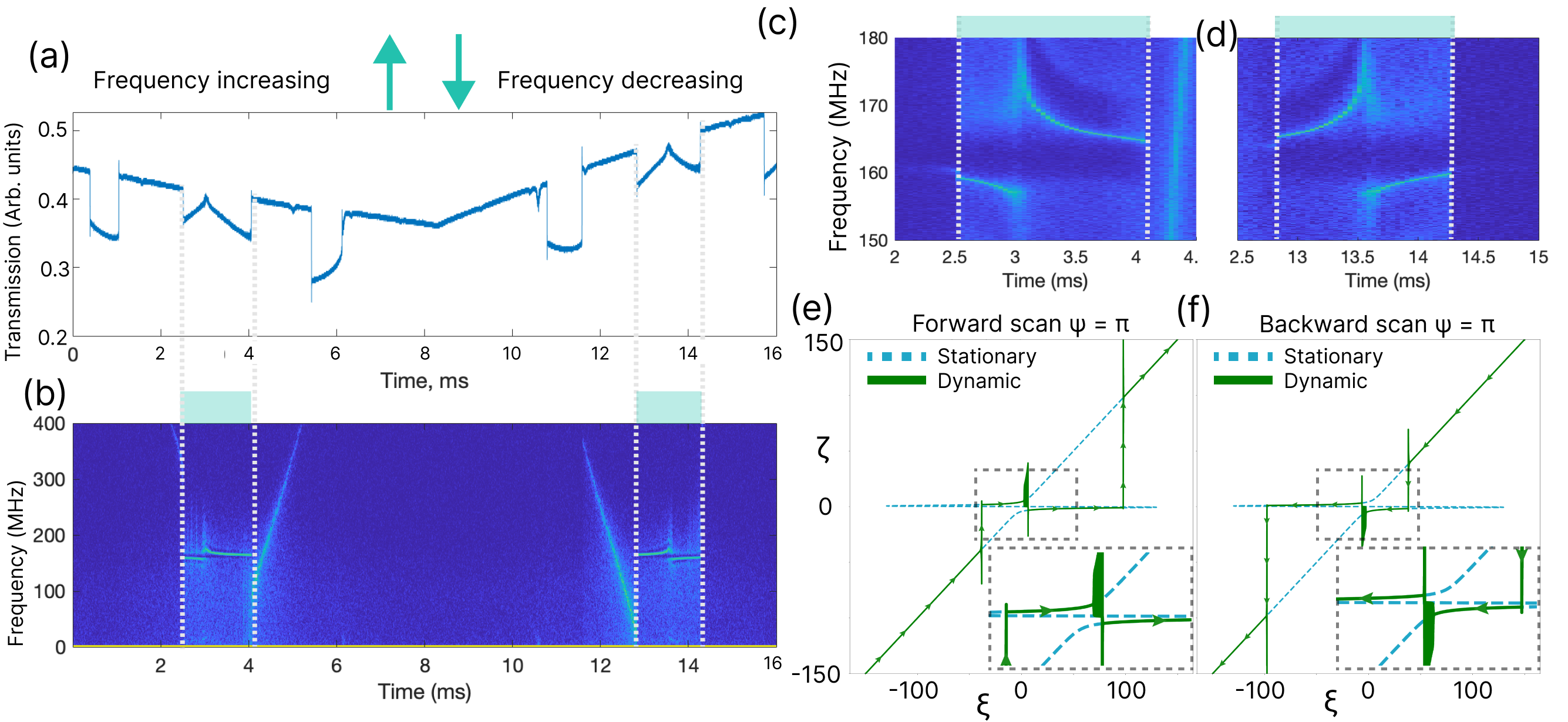}
\caption{SIL dynamics at locking phase $\psi = \pi$. The transmission signal is presented in the top graph (a). The spectrogram of the beatnote signal with the reference laser is presented in the middle (b). Enlarged areas of the spectrograms  corresponding to locking ranges are presented in the panels (c, d). The modeling results are presented in the bottom panels (e, f). In the inset enlarged areas of the transmission curves are presented. Analytical stationary solutions are plotted with blue dashed line and dynamic modeling with solid green line.
}

\label{fig:PhaseComparisonPi}

\end{figure}

The spectrograms of the transmission spectrum for optimal locking phase ($\psi = 0$) and non-optimal locking phase ($\psi = \pi$) are presented in Figs. \ref{fig:PhaseComparison0} and \ref{fig:PhaseComparisonPi} as the most demonstrative ones to analyse the SIL changes during locking phase changes. Laser diode current is monotonously increased from 0 to 8 ms and decreased from 8 to 16 ms, which corresponds to frequency decreasing and frequency increasing consequently. If there were no SIL the laser frequency would pass through zero frequency, where it is equal to heterodyne frequency, and then increase until the turning point at 8 ms. That is because we calculate the absolute value of the frequency with Fourier transform. In Fig.~\ref{fig:PhaseComparisonPi} the tuning range is larger than in Fig.~\ref{fig:PhaseComparison0}.
The SIL regions are marked with light green stripes above the panels and circumscribe with dashed lines. Transmission spectra are presented in the top panels (a) of Figs. \ref{fig:PhaseComparison0} and \ref{fig:PhaseComparisonPi}. The consequent spectrograms are presented in the middle plots (b) and enlarged areas of the SIL regime are presented in the bottom panels (c), (d). 
Almost equal widths of the locked regions for the laser frequency increasing and decreasing is supposed to be an indicator of $\psi = 0$ [see Fig.~\ref{fig:PhaseComparison0}]. A high-order mode is also excited at 11 ms in the (a) panel. It is pronounced in one direction of the frequency scan because of the SIL which takes place asymmetrically considering microresonator eigenfrequency, see Fig.~\ref{fig:PhaseComparison0} (e, f). It can be noticed in the spectrogram in the panel (b) at 11 ms as a small plateau.
The case of $\psi = \pi$ is also symmetric regarding the current sweep however, the frequency change in the locking range is a lot higher, indicating lower stabilization coefficient\cite{Kondratiev:17}. Furthermore, the splitting of the tuning curve can occur in this regime [see Fig.~\ref{fig:PhaseComparisonPi}(b,c,d)]. The position of the jump is utterly sensitive to the locking phase. The value of the jump depends on the quality factor, and it disappears in an overcoupled regime when $\kappa_{\rm mc}$ becomes dominant, but persists in the undercoupled regime when the quality factor is still high. This phenomenon can be explained as follows: due to the high Q factor of the microresonator, even weak amplitudes of the incoming waves can be resonantly amplified and backscattered (with a rate about 1/150) to stabilize the laser emission. In other words the feedback from the edge of laser emission line is resonantly reflected and initiates the transition. 
So the finite laser emission linewidth causes the spontaneous transition of the laser emission frequency to the branch with higher stabilization coefficient [see Figs.~\ref{fig:PhaseComparison0},\ref{fig:PhaseComparisonPi}(e, f)] before the turning point. We performed a dynamical modeling of the self-injection locking\cite{Kondratiev:17} and found that such spontaneous transition is possible after certain microresonator quality factor value [see green lines Figs.~\ref{fig:PhaseComparison0},\ref{fig:PhaseComparisonPi}(e, f)]. Similar spontaneous locking was also seen in nonlinear regime in \cite{Kondratiev2020}. We note that laser rate equations automatically generate nonzero linewidth in the modelling due to numerical noise.
As can be seen in Fig.~\ref{fig:PhaseComparisonPi}, the experimental observations (c,d) are in full agreement with the theoretical simulations of SIL dynamics (e, f) and the corresponding splittings for forward and backward scan in the case $\psi = \pi$ are present.

One may use spectrograms to measure the stabilization coefficient according to \eqref{K} as a ratio of the free-running laser frequency change with the injection current $K_{\rm free}$ [e.g. line slope in unlocked region in Fig.~\ref{fig:PhaseComparison0}(b)] and a locked laser frequency change $K_{\rm lock}$ [e.g. line slope in Fig.~\ref{fig:PhaseComparison0}(c),(d)].
The values for $K_{\rm free}$ and $K_{\rm lock}$ for different tunable parameters were obtained from spectrograms. 
We make linear approximation of the frequency tuning of the laser in unlocked regime for more than hundred independent measurements and calculate average value of the $K_{\rm free}\, = \, (44 \pm 4) $   MHz/ms. 
Then, we approximated frequency dependence in locked states with fifth order polynomials and find minimum of the frequency slope that corresponds to  maximum $K_{\rm lock}$, see locked areas in Fig.~\ref{fig:PhaseComparison0}, \ref{fig:PhaseComparisonPi}(c, d). The ratio of the obtained values of $K_{\rm free}$ and $K_{\rm lock}$ allows us to determine the stabilization coefficient $K$, as defined in equation \eqref{K}. 
Thus, an experimental dependence of the stabilization coefficient on the gap (coupling) and locking phase are obtained.
The linewidth of a free-running laser is approximately 2 MHz.
The squared stabilization coefficient maximum is $(57 \pm 5)\cdot10^3$, so according to Eq. \eqref{K2} we expect locked linewidth of $\approx$ 40 Hz, additional information about calculating stabilization coefficient from Eq. \eqref{StabCoeffGeneral} is presented in supplementary materials.

\begin{figure*}[htbp!]
\centering
\includegraphics[width= \linewidth]{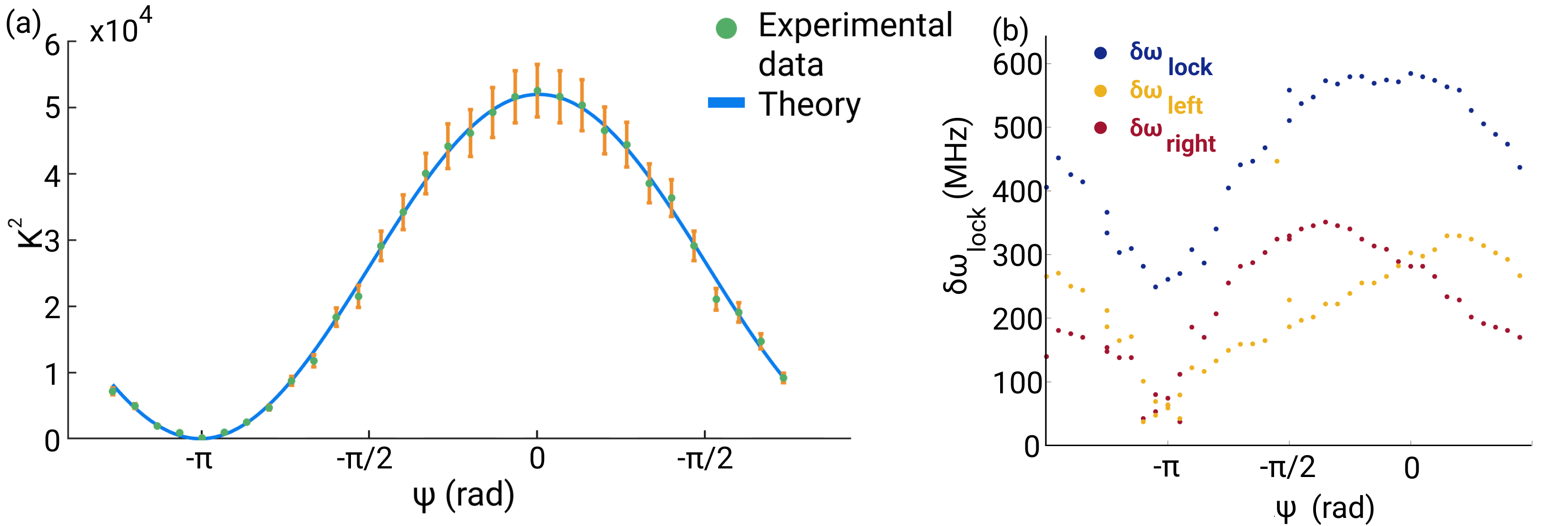}
\caption{(a) The dependence of the stabilization coefficient on the locking phase. The theoretical curve is calculated using Eq. \eqref{StabCoeffCritical} and data presented as dots are obtained from the spectrograms using Eq. \eqref{K} ($\delta\omega_{\rm crit}/(2\pi) = 610 \pm 20$ MHz and $\kappa_{\rm mi}/(2\pi) = 3.73 \pm 0.22$ MHz, which corresponds to the microresonator intrinsic quality factor $Q_{\rm int} = (5.1 \pm 0.3)\cdot10^7$). (b) The dependence of the locking range width on the locking phase.
}
\label{fig:Phase}
\end{figure*}

To analyze how the stabilization coefficient of the SIL changes with the locking phase variation [see Fig.~\ref{fig:Phase}] we measure beatnotes for the spectrograms for different locking phases in the laser diode displacement range of $0.8\ \mu $m. It corresponds to phase rotation for more than $2\pi$ since it changes the optical path for more than $\lambda$. The stabilization coefficient dependence on the locking phase is calculated from the spectrograms according to Eq. \eqref{K} and compared with values obtained using \eqref{StabCoeffCritical} with $\delta\omega_{\rm crit}$ taken from experimental data. The experimental results are in a good agreement with theoretical curve based on Eq. \eqref{StabCoeffCritical} shown in Fig.~\ref{fig:Phase}(a).

The locking range is also dependent on the locking phase [see Fig.~\ref{fig:Phase}(b)]. We measure locking range that is sum of the locking ranges during frequency increasing $\delta\omega_{\rm right}$ and decreasing $\delta\omega_{\rm left}$: $\delta\omega_{\rm lock} = \delta\omega_{\rm right}+\delta\omega_{\rm left}$. It can be seen that the period of the stabilization coefficient $K$ and locking width from the locking phase coincide. The maximum of K is localized near the $ \psi = 0 $ phase, where the widths of the locking ranges at both directions of the frequency scan become equal. The nonsmoothness of the curves can come from the fact that the locking width is not determine well due to spontaneous locking.

\section {Coupling optimization}

Another important controllable parameter of the SIL is the coupling rate. The simple relationships that describe the dependence of the SIL laser spectral characteristics on the distance between the coupling element and the WGM microresonator is obtained in Section~II \eqref{StabCoeffSimplified}, \eqref{kappa}. We also developed an original setup [see Fig.~\ref{fig:Setup}(a)] and performed experimental verification of the obtained theoretical results.

We use spectrograms to measure dependencies of the stabilization coefficient and locking width on a gap between the coupling prism and a microresonator. Locked areas expand until critical coupling and get narrower during overcoupling. We measure locking width and stabilization coefficient for the different values of the gap with 27 nm step. The dependence of SIL parameters on the gap are measured for the optimal locking phase near $\psi = 0$, and its value remains constant during loading as the gap is not included into optical path.

\begin{figure}[htbp!]
\centering
\includegraphics[width=0.65\linewidth]{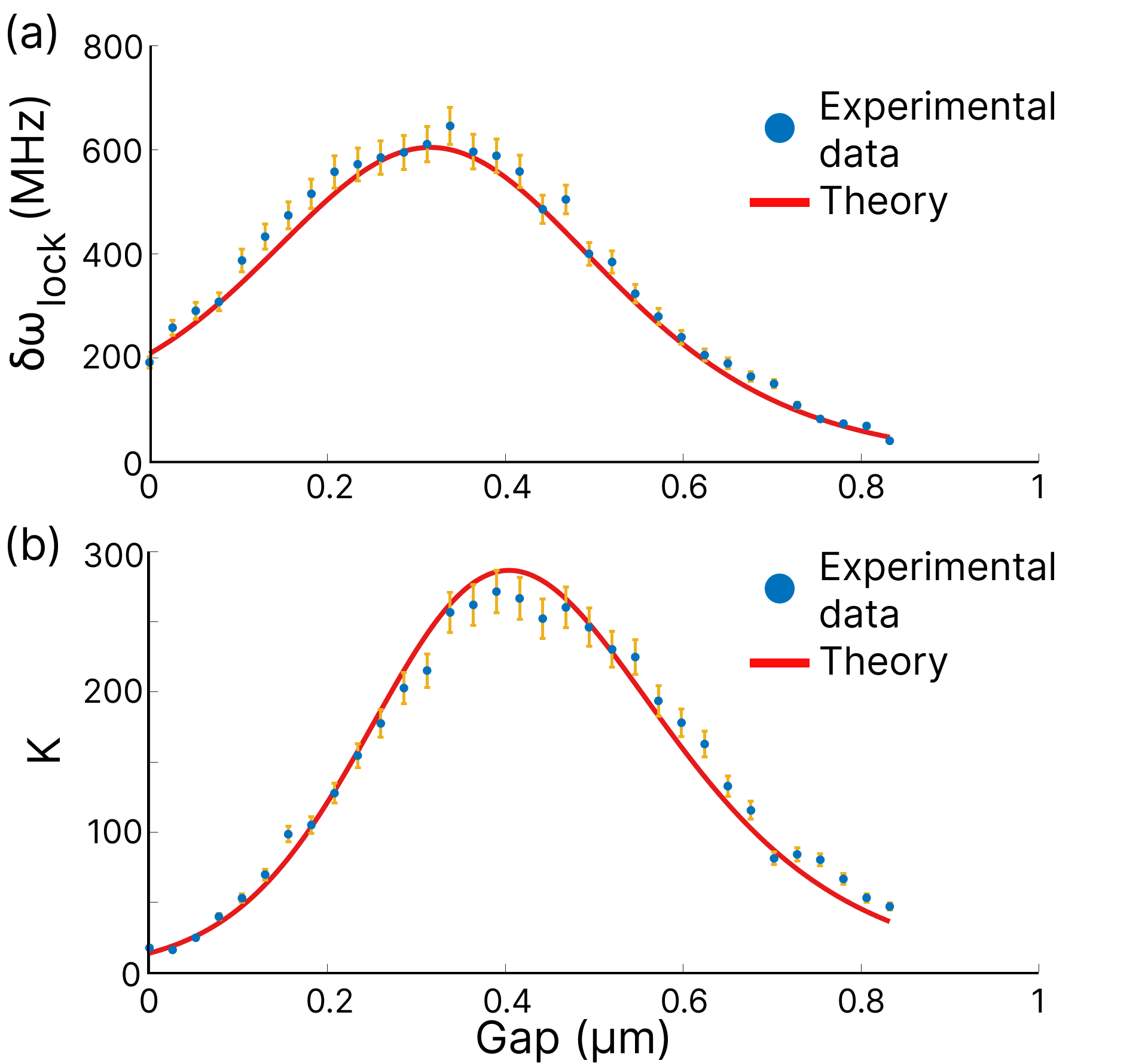}
\caption{The dependence of the locking range (a) and the stabilization coefficient (b) on the gap between the coupling element and the microresonator. The experimental data are plotted with dots and theoretical estimations according method from \cite{shitikov2020microresonator} and Eq. \eqref{StabCoeffSimplified} are plotted with solid lines. Approximation parameters are close and overlapped with error deviations $\kappa_{\rm mi}/(2\pi) = (3.73\pm0.22)\cdot10^6$ Hz for the locking width and $\kappa_{\rm mi}/(2\pi) = (3.82 \pm0.09)\cdot10^6$ Hz for the stabilization coefficient. One may see the extremum of the stabilization coefficient shifted to the undercoupled regime, which is exact consequence of Eq. \eqref{StabCoeffSimplified}. 
}
\label{Shir}
\end{figure}

The dependencies of the locking range width and stabilization coefficient on the gap between coupler and microresonator are presented in Fig.~\ref{Shir} in the top and bottom panels consequently. We approximate the locking range dependence on the gap using the method described in \cite{shitikov2020microresonator}. The approximation parameters were determined as follows: vertical index of the mode $p = 2$, $\kappa_{\rm mi}/(2\pi) = 3.73\pm0.22$ MHz $(Q_{\rm int} = (0.51 \pm 0.06) \cdot 10^8)$ and locking range at critical coupling $\delta\omega_{\rm crit}/(2\pi) = 610 \pm 20$ MHz. The maximum of the locking range corresponds to the critical coupling. 

The experimental values of the stabilization coefficient obtained from the spectrograms were approximated by a theoretical curve (Eq. \eqref{StabCoeffSimplified}). The approximation parameters were determined as follows: vertical index of the mode $p=2$, $\kappa_{\rm mi}/(2\pi) = 3.82 \pm0.09$ MHz $(Q_{\rm int} = (0.50 \pm 0.01) \cdot 10^8)$. 
The value of $\kappa_{\rm mi}$ was used for the calculations of the separately measured stabilization coefficient dependence on the locking phase in Fig.~\ref{fig:Phase}(a) using Eq. \eqref{StabCoeffCritical}. The locking range presented in Fig.~\ref{fig:Phase}(b) at the optimal phase coincides with the locking range at critical coupling in Fig.~\ref{Shir}(a) that indicates optimal phase during the experiment with coupling variation. Finally, the maximum stabilization coefficient from Fig.~\ref{fig:Phase}(a) 239 $\pm$ 15, and maximum $K$ from Fig.~\ref{Shir}(b) exceeds 270, but at the critical coupling it is from 215 to 256, that is perfect agreement. Such a close relationship between the parameters obtained independently in various experiments indicates a high reliability of the theory.  

It is worth discussing that maximum of the stabilization coefficient does not coincide with critical coupling, that is completely according to theoretical estimations. It was shown in \cite{Galiev20} that the optimal coupling condition for low $\beta$ tends to $\eta=1/3$. The value of the stabilization coefficient is decreased by 20\% at critical coupling and since the laser linewidth is inverse proportional to the stabilization coefficient squared it seems even more important. Also, consequently, any technical deviations, for example, of the locking phase or injection current, will be suppressed more effectively in the undercoupled regime than in critically coupled one that will influence not only the instantaneous linewidth but the relatively long-term stability.
In the case of the perfect mode matching the output from the prism for critical coupling is equal to zero. So, the obtained optimal value is appropriate for applications where the uncoupled light is used.

\section{SIL laser characterization}


\begin{figure}[htbp!]
\centering
\includegraphics[width=0.8\linewidth]{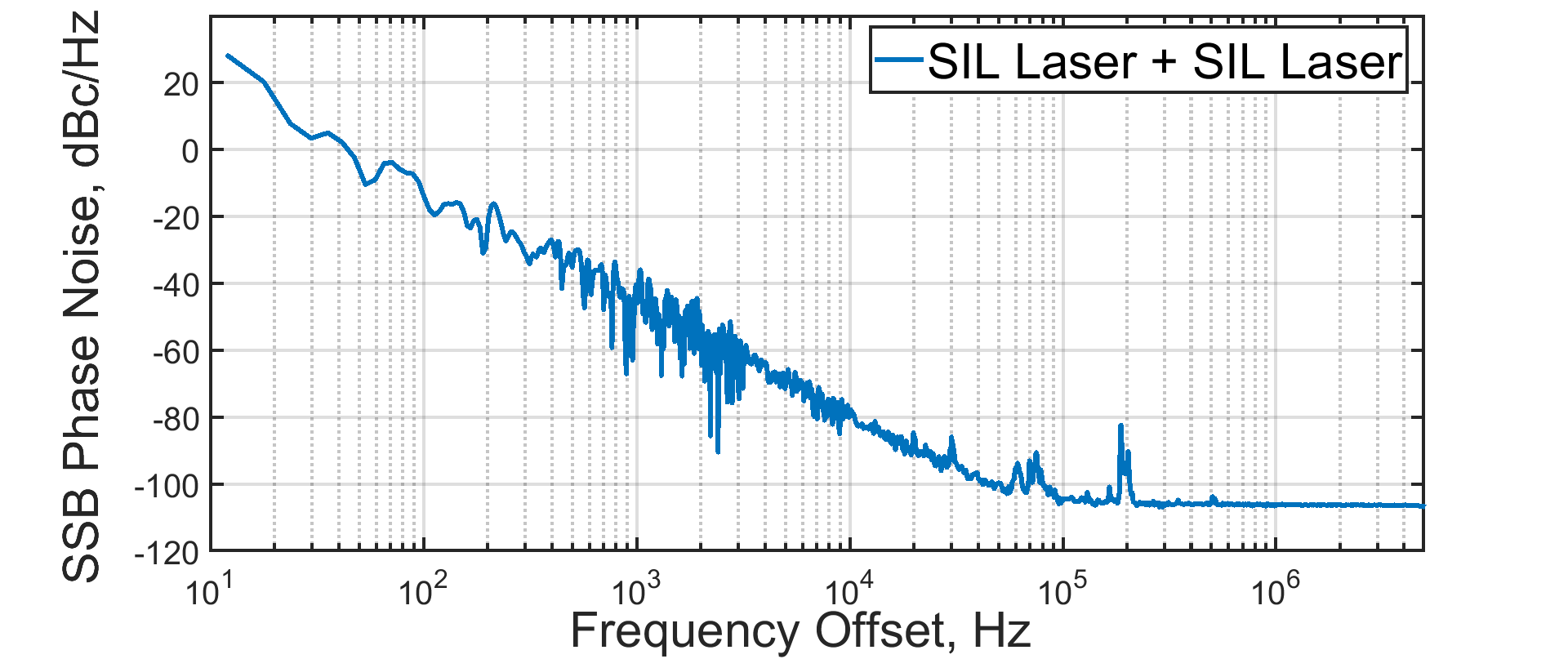}
\includegraphics[width=0.8\linewidth]{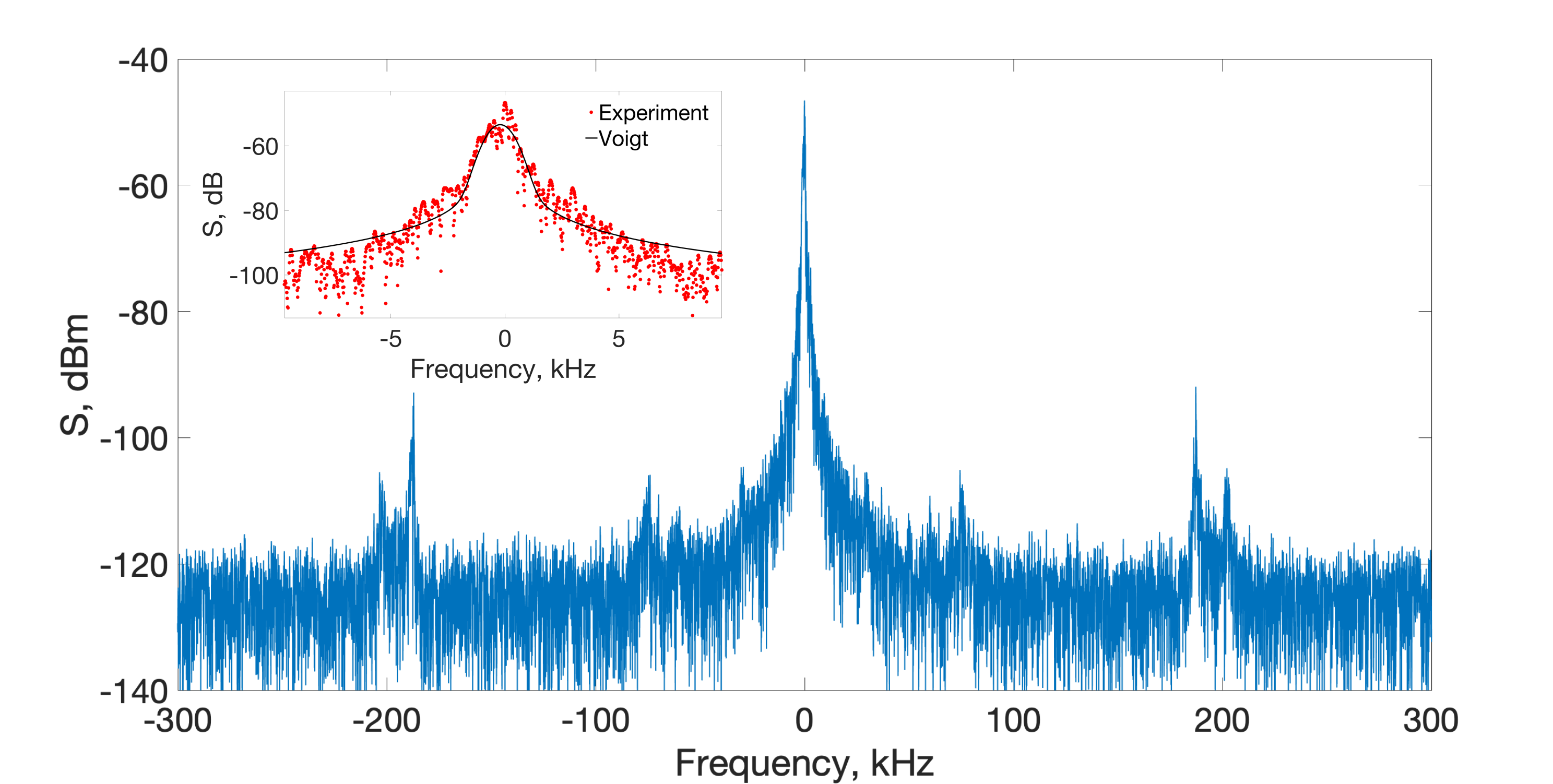}
\caption{Single sideband spectral density of the relative phase noise of the two equivalent SIL laser diodes is presented in the top panel. The beatnote spectrum is presented below. The enlarged area of the peak with Voigt fit is presented in the inset, Lorentzian part is 25 Hz and Gaussian part is 540 Hz. Linewidth measurements performed with 100-Hz resolution bandwidth.
}
\label{fig:PhaseNoise}
\end{figure}

An experiment was carried out to measure the beatnote signal of two SIL lasers stabilized with high-Q WGM microresonators. The coupling was set up close to the critical. The injection current was set up close to the laser generation threshold in both cases to avoid nonlinearities and reduce thermal frequency deviations. Varying injection current and laser diode temperature we found WGMs which eigenfrequencies were different from each other by 3 GHz and both modes simultaneously provided effective stabilization. The loaded quality factors of the modes were approximately $10^8$. Single sideband spectral density (SSB) of the beatnote phase noise is presented in the top panel of Fig.~\ref{fig:PhaseNoise}. At the frequency offset of $5\cdot10^4$ Hz phase noise level corresponds to instantaneous linewidth about 1 Hz. At higher frequency offsets the noise level of the measuring equipment become dominant, see the plateau from $10^5$ Hz.

The beatnote signal is presented in the bottom panel of Fig.~\ref {fig:PhaseNoise}. The approximation of the beatnote is presented in the inset. Linewidth measurements performed with 100-Hz resolution bandwidth. To analyze the beat data, we use an approximation with the Voigt profile (black curves in the panels). The Voigt profile allows us to estimate of the contribution of the white noise (responsible for the Lorentzian linewidth) and the flicker noise impact (Gaussian linewidth). The Gaussian part of the approximation is equal to 540 Hz and Lorentzian part is 25 Hz.

\section {Linear frequency modulation}

\begin{figure}[htbp!]
\centering
\includegraphics[width=0.65\linewidth]{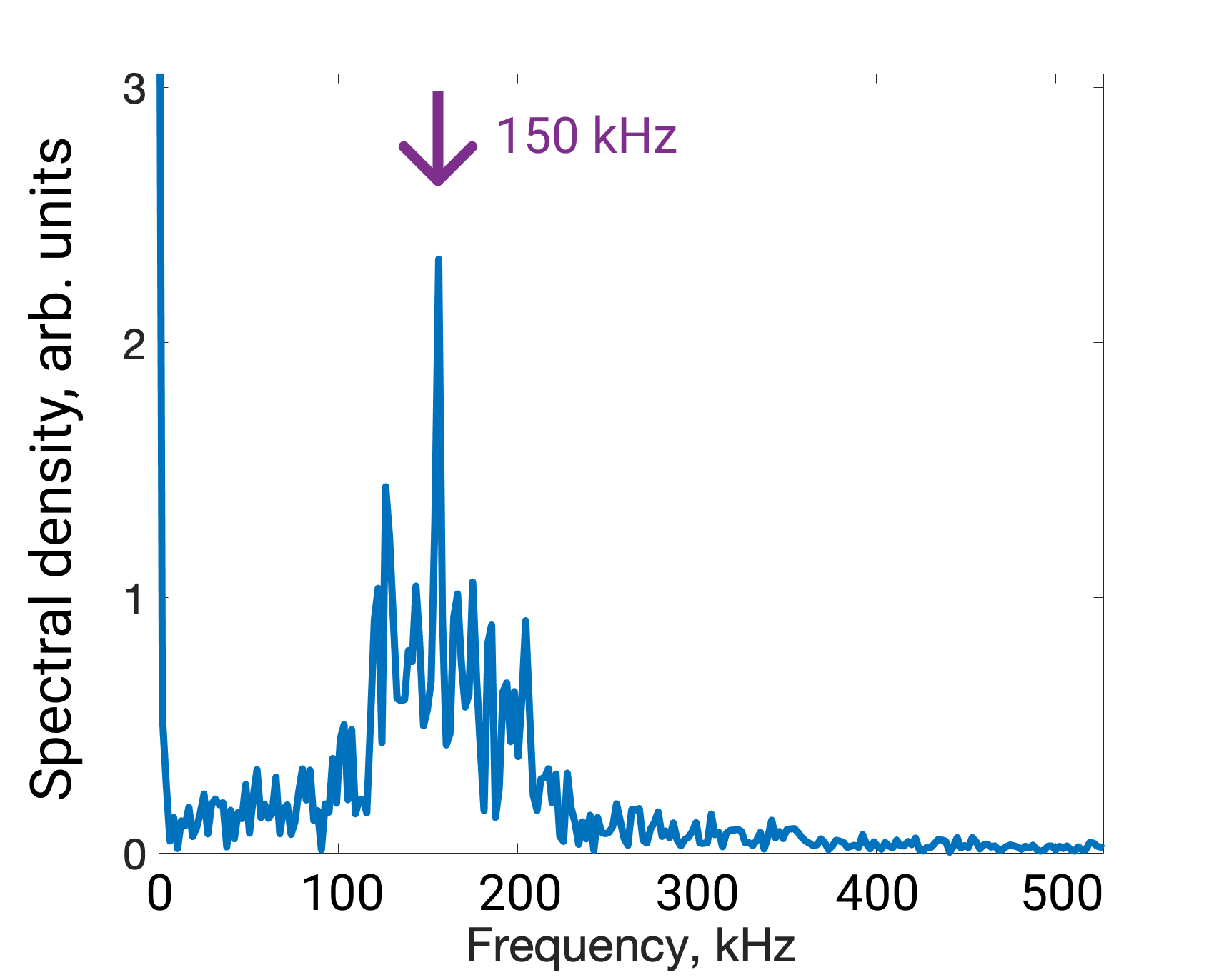}
\caption{Spectrum of the signal of the self-beatnote with the delayed light during coherent optical ranging. The laser was modulated inside SIL regime with the period of 1 ms. 
The peak, locateded at 150 kHz, corresponds to amplitude of the frequency modulation of 1.5 MHz given the 10 km delay line, that is in a good agreement with frequency changing inside SIL [see Fig.~\ref{fig:PhaseComparison0}(c,d)]. 
}
\label{fig:Modulation}
\end{figure}

For many important purposes it is necessary to realize linear frequency tuning. The ability to manipulate the frequency linearly by controlling the injection current of a laser diode was shown for conventional diode lasers. However, due to instabilities, special tricks or additional equipment are required to linearize the tuning, for example, machine learning can be used for this purpose to compensate the nonlinear frequency response for a particular laser diode by restructuring the injection current \cite{zhang2019laser}.

In Fig.~\ref{fig:PhaseComparison0}(c,d) one can find the areas where the laser frequency changes linearly with the injection current in the SIL regime. 
Moreover, these sections are simultaneously the regions of the highest stabilization coefficient. The deviation of the frequency changes in this case is 1-10 MHz and depends on the position on the tuning curve, locking phase, and coupling.

We observed linear frequency modulation with frequencies from 150 Hz to 200 kHz. The resonator was critically coupled, and the locking phase was close to optimal one. The maximum amplitude of the injection current modulation depends on the locking range width, the wide range is the higher injection current modulation inside SIL can be achieved. The injection current modulation is applied by means of RF-generator. 

We conducted experiment to demonstrate conceptually a FMCW LIDAR in the SIL regime. The laser frequency was modulated linearly inside the locking range [e.g. with amplitude 1.5 MHz, see Fig.~\ref{fig:PhaseComparison0}(c,d)] with a period of 1 ms. Note that the frequency increases and decreases with the rate of 3 MHz/ms for that amplitude during the period. The laser beam was split into two parts, one was propagating through a 10 km delay line and the other -- only through a polarization controller. The spectrum of the self-heterodyne signal is shown in Fig.~\ref{fig:Modulation}. To calculate the spectrum, we took the signal sections minus 100 microseconds near the turning point which appear when the frequency tuning changes direction. The length of the optical fiber is $l=10$ km which corresponds to the time delay of $\tau = l n_{\rm eff}/c \approx 50 \ \mu $s, where $n_{\rm eff}$ is effective refractive index of the optical fiber. The frequency difference between the beams caused by the time delay is expected to be 150 kHz as a delay multiplied on the frequency changing rate, which is in good correspondence with the measured spectrum.
If we will turn off the self-injection locking the self-heterodyne signal will disappear.

\section {Velocity measurements}
Diode lasers are widely available efficient sources of laser radiation, however, for measuring velocities of the order of $\mu $m/s, the use of these lasers is limited due to the linewidth of the order of 1 MHz, as discussed, for example, in \cite{donati2012}, as well as in \cite{ lu2012}. SIL lasers seem to be ideal candidates for measuring low velocities, e.g. via Doppler shift measurement, due to achievable ultra-narrow linewidths.

The experimental setup for the measurement of the Doppler frequency shift of the reflected signal of the SIL laser is presented in Fig.~\ref{fig:Setup}(e). 
\begin{figure}[htbp!]
\centering
\includegraphics[width=0.9\linewidth]{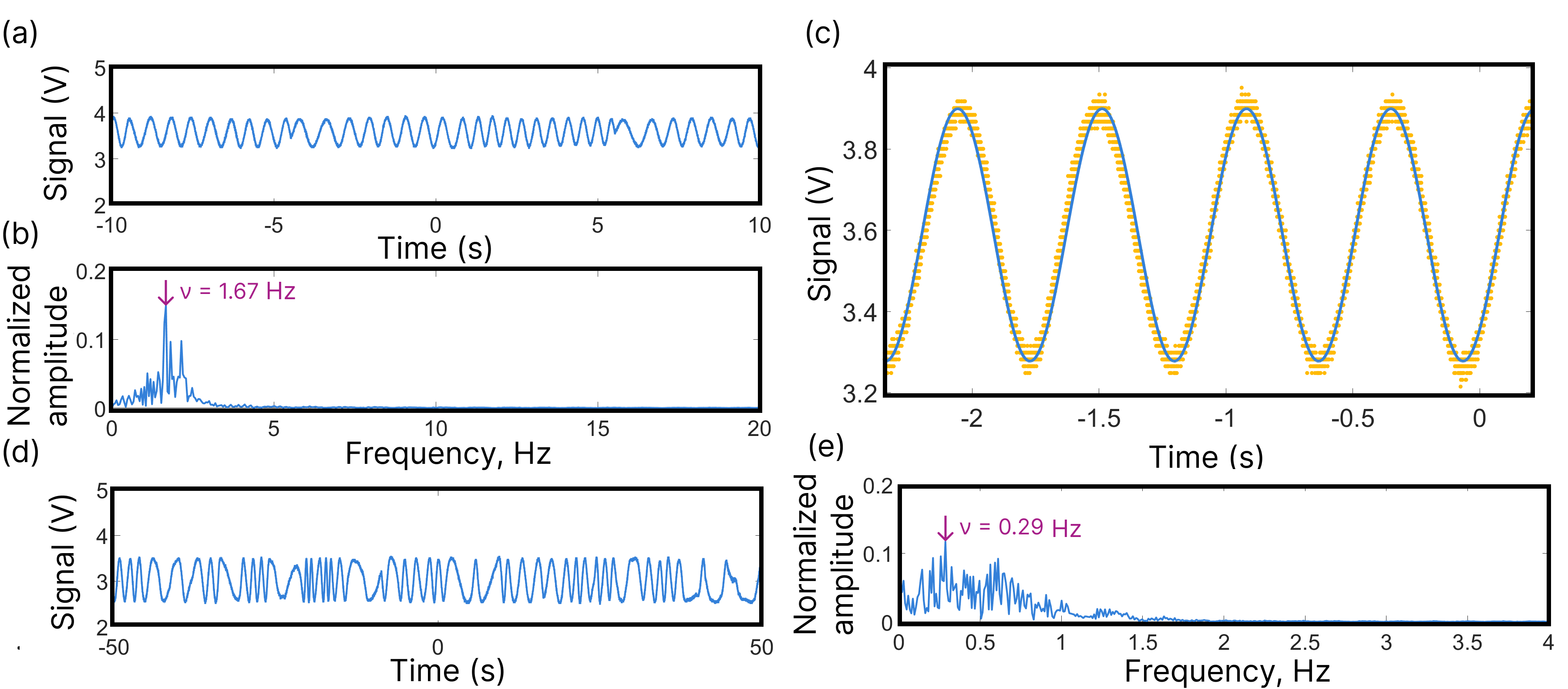}
\caption{Reflected signal from the moving target. (a) The signal in case of velocity of the target of 1.4 $\mu $m/s over 20 s measurement time. (b) Fourier transform of signal from (a). The maximum is observed at 1.67 Hz which is close to expected 1.75 Hz. (c) The sector of the signal fitted with 1.75 Hz. (d) The signal in case of velocity of the target 0.27 $\mu $m/s over 100 s. The signal is noisier than for 1.4 $\mu $m/s. (e) Fourier transform of signal from (d). The maximum is observed at 0.29 Hz which is close to expected 0.35 Hz. 
}
\label{fig:VelocityMeas}
\end{figure}
Diffuse reflective plate was used as a target. It was placed in the focus of a converging lens. The plate was fixed on the translation stage with a PZT. Ramp voltage with amplitude of 4 V that corresponds to 6.8 $\mu$m displacement and frequency of tens of mHz was applied to the PZT. As can be seen from the signal, in the vicinity of the turning point, the frequency of the sinusoidal changes [see Fig.~\ref{fig:VelocityMeas}(a)]. Fourier transform of this signal gives us maximum at 1.67 Hz [see Fig.~\ref{fig:VelocityMeas}(b)]. The frequency of the signal is close to expected (1.75 Hz for velocity of 1.4 $\mu$m/s, if we extract the part corresponding to the uniform motion of the surface between the turning points. We extracted the sector from (a) and fitted it with sinusoidal function with 1.75 Hz [see Fig.~\ref{fig:VelocityMeas}(c)]. It is obvious that if we turn off SIL by receding the microresonator the frequency cannot be measured in that way due to high level of phase noise. 

The lowest velocity we observed was 0.27 $\mu $m/s [see the signal in Fig.~\ref{fig:VelocityMeas}(d)]. The maximum of the Fourier transform was located at 0.29 Hz instead of 0.35 Hz expected, that corresponds to 0.22 $\mu $m/s instead of 0.27 $\mu $m/s. This variance can be caused by roughness of the target which can be pronounced at this level of measured velocities. Nevertheless, demonstrated accuracy conceptually proves the effectiveness of SIL lasers for such sort of measurements.

\section{Conclusion}
We investigate in detail such aspects of the self-injection locking phenomena as dependence of the spectral characteristics on the locking phase and loading. 
We have verified experimentally the dependence of the stabilization coefficient on the locking phase, found the optimum and showed that the resulting linewidth remains narrow in a wide range of phases. We have shown that one can vary the locking phase within the limits defined by the desired linewidth and thus tune the laser frequency in a range of several MHz. Due to the purely optical essence of SIL the tuning can be very fast and, in our case, limited by the PZT drive. The phenomenon of spontaneous locking was observed and its bad influence on the locking at $\pi$-phase was revealed.
We demonstrate that the locking phase does not change with loading. It was demonstrated experimentally that the maximum of the stabilization coefficient appears in the undercoupled regime and does not coincide with critical coupling.
The SIL lasers used in experiments demonstrate superior stability, its phase noise single sideband spectral density was measured by heterodyning of two equivalent laser sources and 1 Hz instantaneous linewidth was demonstrated analyzing phase noise spectral density and 25 Hz linewidth by beatnote approximation with Lorentzian profile. 
As an example of possible applications, we demonstrated linear frequency modulation by the injection current manipulation inside SIL regime. The repetition rate was up to 200 kHz. This modulation was used to measure the length of a $\simeq 10 $ km delay line. The SIL laser source was also used for sub-$\mu $m velocity measurements. These experiments demonstrate new ways of the SIL scheme optimization and implementation in a fashion of the perspective of application SIL lasers for long distance measurements, space debris monitoring and controlling, distant object classification and other applications.



\begin{backmatter}
\bmsection{Funding}
RQC team was supported by the Russian Foundation for Basic Research (Project 19-29-06104mk).

\bmsection{Acknowledgments}
AES and VEL acknowledge the Foundation for the Advancement
of Theoretical Physics and Mathematics “BASIS” for personal support.

\bmsection{Disclosures}
The authors declare no conflicts of interest.

\bmsection{Data availability} Data underlying the results presented in this paper are not publicly available at this time but may be obtained from the authors upon reasonable request.

\bmsection{Supplemental document}
See Supplement 1 for supporting content.
\bigskip

\end{backmatter}


\bibliography{sample}






\end{document}